\documentclass[aps,prl,twocolumn,nofootinbib,superscriptaddress,floatfix]{revtex4-2}

\usepackage[T1]{fontenc}
\usepackage[utf8]{inputenc}
\usepackage{bm}
\usepackage[colorlinks=true,linkcolor=blue,citecolor=blue,urlcolor=blue]{hyperref}
\usepackage{microtype}
\usepackage{physics}
\usepackage{slashed}
\usepackage{xcolor}
\usepackage{comment}
\usepackage{multirow}
\usepackage{graphicx}
\usepackage{graphics}
\usepackage{float}
\usepackage{cancel}
\usepackage{soul}
\usepackage{cases}
\usepackage{array}
\usepackage{mathtools}  
\usepackage{amsfonts}
\usepackage{hyperref}
\usepackage{amsmath}
\usepackage{amssymb}
\usepackage{accents}
\usepackage{tcolorbox}
\usepackage{tikz}
\usepackage{tikz-3dplot}

\usepackage{amsmath,amssymb,bbm}
\usepackage{booktabs,longtable,array}
\usetikzlibrary{calc}

\usepackage[table]{xcolor}
\usepackage{tcolorbox}
\tcbuselibrary{skins, breakable}
\usepackage{float}

\newcommand{\Vol}{\operatorname{Vol}}

\newcommand{\Block}{\mathcal{K}}
\begin{document}

\title{Cosmological Correlators from Generating Potentials and Projective Geometry}

\author{Aswini Bala}
\author{Sachin Jain}
\author{Deep Mazumdar}

\affiliation{
Indian Institute of Science Education and Research, Pune, India\\
\textit{aswini.bala@students.iiserpune.ac.in},
\textit{sachin.jain@iiserpune.ac.in},
\textit{deepkamal.mazumdar@students.iiserpune.ac.in}
}
\date{\today}

\begin{abstract}
We introduce novel scalar generating potentials to compute three and four point cosmological correlators in orthogonal  Grassmannian space. These correlators are obtained by applying appropriate directional derivatives on these potentials, combined with helicity-dependent minors. We emphasize that, this is not a spin-raising analog, but a much more powerful construction. The four point generating potential is associated with the geometry of a projective triangle. First-order directional derivatives generate gluon correlator, while the Hessian of the same scalar potential produces the four-graviton correlators, thus realizing a differential double copy. Thus, double copy is a natural consequence of the underlying scalar potential as the origin of the gluon and graviton four point correlators. This framework extends naturally to any spin-\(s\). We exhibit a geometric origin for the gluon colour identities: the projective homogeneity of the scalar potential implies Kleiss–Kuijf identity, while the kinematic Jacobi identity follows as a consequence of the closure of the triangle under directional derivatives. We also uncover a connection between the BCFW bridge deformation of the scalar potential and directional derivatives that produces the spin-s correlator. Thus the generating potential provides a universal geometrical interpretation for any spin-\(s\) four point Grassmannian correlator, unlike \cite{Bala:2026lvw} where geometric interpretation was only shown for the gluons.
\end{abstract}

\maketitle

\vspace{-12.5pt}\section{Introduction}\vspace{-10pt}
Modern amplitude program has led to remarkable success in the study of gluon and graviton amplitudes. These include {Parke-Taylor formula} \cite{Parke:1986gb}, {BCFW recursion} \cite{Britto:2004ap,Britto:2005fq}, {twistors} \cite{Witten:2003nn,Arkani-Hamed:2009hub}, {Grassmannian} \cite{ArkaniHamed:2009Duality,Arkani-Hamed:2012zlh} and the {amplituhedron} \cite{Arkani-Hamed:2013jha}. 
% The Parke-Taylor formula condenses tree-level MHV gluon amplitudes in flat space into a single
% expression \cite{Parke:1986gb}.
However, an equally compact Parke-Taylor like expression for spinning correlators in (A)dS remains elusive. Finding one remains an important achievement for cosmological correlators.

Conventionally, the computation of (A)dS correlators is done in momentum-space and spinor helicity variables \cite{Maldacena:2002vr,Raju:2010by,raju2011recursion,Maldacena:2011nz,McFadden:2011kk,Raju:2012zr,Raju:2012zs,Ghosh:2014kba,Kundu:2014gxa,Kundu:2015xta,Arkani-Hamed:2015bza,Arkani-Hamed:2018kmz,Albayrak:2019asr,Baumann:2019oyu,Baumann:2020dch,Baumann:2022jpr,Bonifacio:2022vwa,Chowdhury:2025ohm,Albayrak:2026auu}. However, they involve complicated time/radial integrals and intricate energy singularities. Twistor variables organize tensor structures for spinning correlators \cite{Baumann:2024ttn,Bala:2025gmz,Bala:2025jbh,Bala:2025qxr,Mazumdar:2025egx,CarrilloGonzalez:2025qjk}, but studying even four point correlators has been a computationally challenging problem \cite{Ansari:2025fvi}. There has been recent progress in generalizing correlators to $n$-points \cite{Albayrak:2018tam,Albayrak:2019yve,Armstrong:2022mfr,Chu:2023pea,Mei:2024abu,Mei:2024sqz,Gomez:2026yno}, but an analog to Parke-Taylor formula still remains to be understood.

A promising approach to study (A)dS correlators was introduced, analogous to the flat space program, using the Grassmannian formulation \cite{Arundine:2026fbr}. The cosmological Grassmannian packages the kinematics compactly
\cite{Arundine:2026fbr,De:2026shn,Bala:2026hdm,Bala:2026bdx,Huang:2026tsh,Arundine:2026myr,bala2026color}, and a BCFW-like bridge enables the algebraic reconstruction of four points functions of any spin \cite{Bala:2026lvw}. An interesting geometric description for four point Yang-Mills correlators has been presented in \cite{Bala:2026lvw, Arundine:2026qqg}. While a systematic extension beyond four points remains a challenging endeavor \cite{XYZ:2026new4}, the striking simplicity of Grassmannian four point functions encourages us to explore the possibility of a Parke-Taylor-like formula for the higher point correlators.

In this paper, we introduce a novel framework of scalar generating potentials, defined independently of any particular spinning correlator. At three points, its normalized derivatives generate helicity-stripped equal-spin correlators for any spin. For four points, the generating potential is related to the canonical form of a projective triangle. %whose canonical-form density is $e^{2\Phi_4}$.
Derivatives of this single potential yield gluon, graviton, and higher-spin exchange correlators, giving them a common geometric origin. Given the simplicity of the four point generating potential, it is highly suggestive that the higher point correlators would also be generated by such a simple potential. We then derive the gluon kinematic Jacobi identity from the closure of the projective triangle and obtain the Kleiss--Kuijf relation from projective homogeneity. This provides a geometric root to the color identities. The geometric construction leads to a differential double copy for gravitons and a $s^\textrm{th}$ differential copy spin-$s$. %for exchange numerators.
Our results exhibit a pattern, which suggests that a scalar construction should be an underlying structure for any double copy relation.

We finally establish a connection between the generating potential and the BCFW bridge construction \cite{Bala:2026lvw}.
We thus propose the generating potential as a candidate organizing object at general multiplicity.
%; its higher-point form and physical origin remain to be established.
% 1. Park taylor formula for (A)dS is not there
% 2. calcs in mom space, SH (twistor) difficult
% 3. Grassmannian+BCFW make a strong candidate for calcs
% 4. 5pt still seems complicated
% 5. intersting geometric interpretation at 4pt found
% 6. Generalizing geometry would be great... one mightt belive this would be significantly difficult and challenging'
% 7. we introduce indepenedant scalar potential... simpole, elegent, geometric interpretation, corr fxn as derivatives of seed, shared origin of 4pt geometry for gluon, graviton and HS
% 8. belive this is the object to obtain n-point
% 9. a short para on what has been done in this paper

\vspace{-12.5pt}\section{Orthogonal Grassmannian}\vspace{-10pt}
Consider the Grassmannian setup for CFT$_3$ correlators \cite{Arundine:2026fbr}, where the $n$-point correlator is an integral over the orthogonal Grassmannian OGr$(n,2n)$
\begin{align}\label{npoint}
\psi_n(\Lambda)=\int \frac{d^{n\cross2n}C}{\Vol GL(n)}\;\delta(C\cdot Q\cdot C^T)\delta(C\cdot \Lambda) F_n(C),
\end{align}
where $Q$ is the metric on the kinematic space
\begin{align}
   Q=\begin{pmatrix}
       0& \mathbf 1_{n\times n} \\
       \mathbf 1_{n\times n}& 0
   \end{pmatrix} .
\end{align}
The columns of the matrix $C$ are labeled as $C\in \{\overline{1},\cdots,\overline{n},1,\cdots,n\}$ and $F_n(C)$ is function of the minors $(I_1\cdots I_n)$. We work with the right-branch gauge choice:
$C_n=(1_{n\cross n}\;-c_{n\cross n})$ where $c_{ij}$ is skew-symmetric.

The Grassmannian correlator $F_n(C)$ for two and three points are fixed by demanding correct helicity weight and $GL(n)$ weight. The four point correlators are obtained by then demanding correct flat-space limit and factorization properties too \cite{Arundine:2026fbr}. A more efficient and powerful approach to obtain the four point function uses the BCFW bridge as was shown in \cite{Bala:2026lvw}.

\vspace{-12.5pt}\section{Three Point Generating Potential}\vspace{-10pt}
We define the generating potential for the three-point functions and its normalized derivatives
\begin{align}\label{3potder}
\Phi_3=-\textrm{log}\;\mathcal{K}_3,&\quad D_{3}=-\frac{1}{c_{23}c_{31}}\partial_{c_{12}},
\end{align}
such that $D_3\mathcal{K}_3=-1$. Here $\mathcal{K}_3$ is the unique little-group-neutral object to construct a scalar generating potential
\begin{align}\label{3ptK}
\mathcal{K}_3=(1 \bar 1 2)(\bar 2 3 \bar 3)=c_{12}c_{23}c_{31}.
\end{align}
 The generating potential $\Phi_3$ has a geometric interpretation. It is the logarithm of canonical density. For $c_{ij}>0$
\begin{align}\label{3log}
\Omega_3=&d\textrm{log}\;c_{12}\wedge d\textrm{log}\;c_{23}\wedge d\textrm{log}\;c_{31}=e^{\Phi_3}\;\textrm{d}c_{12}\wedge\textrm{d}c_{23}\wedge\textrm{d}c_{31},
\end{align}
is the canonical 3-form of the positive orthant, whose facets are the coordinate hyperplanes $c_{ij}$ = 0.

\vspace{5pt}\paragraph{\bf{Three Point Functions:}}
The generating potential \eqref{3potder} produces the helicity-stripped three point correlator\footnote{For the case of two points, one can define a scalar potential to be \begin{align*}
    \Phi_2=- \text{log}~\Block_2,~D_2=-\frac{1}{c_{12}}\partial_{c_{12}},
\end{align*}
where $\Block_2=(1\bar1)$ and the generating potential can be used to write a canonical form $\Omega_2= e^{\Phi_2} dc_{12}$. The successive action of normalized derivatives on the potential,
\begin{align*}
f_2^{(s)}=\frac{\Block_2}{(s-1)!}\;(D_2)^s\;\Phi_2=\bigg(\frac{1}{\mathcal{K}_2}\bigg)^{2s-1},
\end{align*}
which gives \[ F_2^{++}(C) = \frac{(\bar1\bar2)^{2s}}{\Block_2^{2s-1}}.\] } 
of equal spin-s, via the successive action of normalized derivatives
\begin{align}\label{fns}
f_3^{(s)}=\frac{1}{(s-1)!}\;D_3^s\;\Phi_3=\bigg(\frac{1}{\mathcal{K}_3}\bigg)^s,
\end{align}
which supplemented by the helicity minor $\mathcal{H}_3$, results in the full three point correlator for \cite{Arundine:2026fbr}
\begin{align}\label{3ptcorr}
F_3^{(s)}(C)=(\mathcal{H}_3)^s\;f_3^{(s)}.
\end{align}
The generating potential renders a geometric picture to helicity-stripped correlators. The gluon correlator is the slope of generating potential $\Phi_3$ along $D_3$, while the graviton correlator is the Hessian of the same potential. The  higher spin correlators are similarly obtained by the higher derivatives of $\Phi_3$. For instance, the gluon correlators for all-plus helicity configuration $\mathcal{H}_3^{+++}=(\bar 1 \bar 2 \bar 3)$ is
\begin{align}\label{3gluon}
F_3^{YM,+++}=\frac{(\mathcal{H}_3^{+++})^2}{\mathcal{K}_3}.
\end{align}

\vspace{-12.5pt}\section{Four Point Generating Potential and Geometry}\vspace{-10pt}
We start by defining the four point Grassmannian Mandelstam variables \cite{Arundine:2026fbr}
\begin{align}
S=c_{13}c_{24}-c_{14}c_{23},\quad& T=c_{13}c_{24}-c_{12}c_{34},\notag\\
U=-c_{14}c_{23}&-c_{12}c_{34}.
\end{align}
Further defining the total and the shifted-energy divisors
\begin{align}
\Sigma=S+T+U,&\quad \Delta_S=-S+T+U,\notag\\
\Delta_T=S-T+U,&\quad \Delta_U=S+T-U,
\end{align}
and normalizing the shifted energy divisors with $\Sigma$, yields
\begin{align}
\widetilde S=\frac{\Delta_S}{\Sigma},\quad \widetilde T=\frac{\Delta_T}{\Sigma},\quad \widetilde U=\frac{\Delta_U}{\Sigma},
\end{align}
where $\widetilde S+\widetilde T+\widetilde U=1$. We introduce a novel four point generating potential%\footnote{Here $\Phi$ organizes the kinematic dependence and is not itself a scalar-field correlator.}
\begin{align}\label{4pot}
\Phi_4=\frac{1}{2}\textrm{log}\bigg(\frac{\Sigma^3}{\Delta_S\Delta_T\Delta_U}\bigg)=-\frac{1}{2}\textrm{log}(\widetilde S\widetilde T\widetilde U),
\end{align}
which is a homogeneous function of degree zero in $(S,T,U)$. We now probe the geometric interpretation of the four point generating potential \eqref{4pot}. 

\vspace{5pt}\vspace{5pt}\paragraph{\bf{Geometry for generating potential:}}

\begin{figure}[H]
\centering
\resizebox{0.425\textwidth}{!}{%
\tdplotsetmaincoords{70}{70}

\begin{tikzpicture}[
    tdplot_main_coords,
    scale=2.0,
    >=stealth,
    line join=round,
    line cap=round
]

% Origin and axes
\coordinate (O) at (0,0,0);
\coordinate (Saxis) at (3.5,0,0);
\coordinate (Taxis) at (0,3.5,0);
\coordinate (Uaxis) at (0,0,3.5);

\draw[->,thick] (O) -- (Saxis)
    node[below right] {$\widetilde S$};
\draw[->,thick] (O) -- (Taxis)
    node[below left] {$\widetilde T$};
\draw[->,thick] (O) -- (Uaxis)
    node[above] {$\widetilde U$};

% Three drawing units represent normalized value 1
\coordinate (A) at (3,0,0);
\coordinate (B) at (0,3,0);
\coordinate (C) at (0,0,3);

% Triangle
\fill[blue!20,opacity=0.75]
    (A) -- (B) -- (C) -- cycle;

% Edges and their equations
\draw[blue!70!black,thick]
    (A) --
    node[midway,sloped,below=2pt]
        {$\widetilde S+\widetilde T=1$}
    (B) --
    node[midway,sloped,above=3pt]
        {$\widetilde T+\widetilde U=1$}
    (C) --
    node[midway,sloped,below=2pt]
        {$\widetilde U+\widetilde S=1$}
    (A);

% Plane label
\path (0.85,0.35,1.80) --
    node[midway,sloped,above=1pt,  text=blue!80!black,font=\scriptsize]
    {$\widetilde S+\widetilde T+\widetilde U=1$}
    (1.0,1.75,0.40);

% Vertices
\fill[blue!70!black] (A) circle (1pt);
\fill[blue!70!black] (B) circle (1pt);
\fill[blue!70!black] (C) circle (1pt);

\node[below left] at (A) {$\widetilde S=1$};
\node[above right]       at (B) {$\widetilde T=1$};
\node[left]       at (C) {$\widetilde U=1$};

% Short channel arrows close to their corresponding edges
\draw[->,black,thick]
    (1.215,1.635,0.15) --
    node[midway,above=1.5pt] {$D_U$}
    (1.635,1.215,0.15);

\draw[->,black,thick]
    (0.12,1.215,1.665) --
    node[midway,below=2pt] {$D_S$}
    (0.12,1.635,1.245);

\draw[->,black,thick]
    (1.635,0.12,1.245) --
    node[midway,above right=2pt] {$D_T$}
    (1.215,0.12,1.665);

\end{tikzpicture}%
}
\caption{Projective triangle associated with the four point
generating potential \eqref{4pot}. Within the positive chamber,
the vertex $\widetilde S=1$ corresponds to the factorization loci
$S=0$. The opposite edge $\widetilde T+\widetilde U=1$ implies the shifted-energy divisor $\Delta_S=0$. The directional derivative $D_S$ in \eqref{4der} generates translation parallel to this edge.}
\label{fig:projective_triangle}
\end{figure}
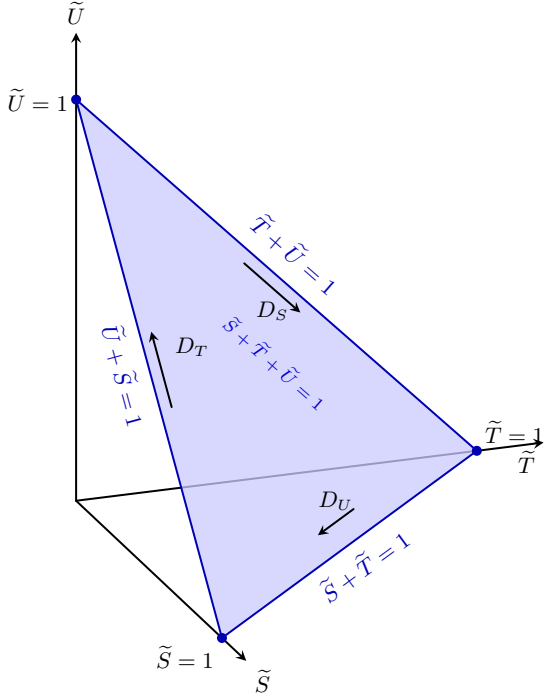

The generating potential \eqref{4pot} is a homogeneous function of degree zero in kinematic variables $\widetilde S, \widetilde T, \widetilde U$. Hence, the geometry associated with the generating potential must depend only on the shape, and not the volume. This makes the geometry projective. The constraint $\widetilde S+\widetilde T+\widetilde U=1$ is a plane, which intersects in the positive orthant $\widetilde S, \widetilde T, \widetilde U>0$ to give a triangle. This leads to a projective triangle (see figure \ref{fig:projective_triangle}), where $\widetilde S, \widetilde T, \widetilde U$ are its barycentric coordinates\footnote{Barycentric coordinates specify the location of a point relative to the vertices of a simplex (here a triangle).}.

The canonical form  of the projective triangle is that of a logarithmic 2-form
\begin{align}\label{4log}
\Omega_4=d\textrm{log}\bigg(\frac{\widetilde S}{\widetilde U}\bigg) \wedge d\textrm{log}\bigg(\frac{\widetilde T}{\widetilde U}\bigg)=\frac{d\widetilde S\wedge d\widetilde T}{\widetilde S \widetilde T \widetilde U}.
\end{align}
This has an elegant connection to the generating potential
\begin{align}\label{4log}
\Omega_4=e^{2\Phi_4}\;d\widetilde S\wedge d\widetilde T.
\end{align}

%%%%%%%%%%%%%%%%%%%%%%%%%%%%%%
Figure \ref{fig:projective_triangle} is the projective triangle associated with the generating potential \eqref{4pot}. The edges {$\widetilde S+\widetilde T=1$, $\widetilde T+\widetilde U=1$ and $\widetilde S+\widetilde U=1$} correspond to the vanishing of the shifted-energy divisors $\Delta_U=0$, $\Delta_S=0$ and $\Delta_T=0$ respectively. The vertices at $\widetilde S=1$, $\widetilde T=1$ and $\widetilde U=1$ correspond to the {factorization} locus $S=0$, $T=0$ and $U=0$ respectively.

We define the normalized directional derivatives\footnote{In the ordinary variable $S,T,U$ one gets $D_U
 =\frac{2}{\Sigma}
   \left(\partial_T-\partial_S\right)=\frac{4}{\Sigma^2}$ and similarly $D_S,D_T.$}
\begin{align}\label{4der}
D_S=\frac{4}{\Sigma^2}(\partial_{\widetilde T}-\partial_{\widetilde U}),&\quad
D_T=\frac{4}{\Sigma^2}(\partial_{\widetilde U}-\partial_{\widetilde S}),\notag\\
D_U=\frac{4}{\Sigma^2}&(\partial_{\widetilde S}-\partial_{\widetilde T}).
\end{align}
The directional derivative $D_S$ generates translation along the edge\footnote{ For example, translation along $\tilde{T}+\tilde{U}=1$, in functional form is represented by  $f(\widetilde T+a,\widetilde U-a)$. The translation can be generated in the differential representation, generated by the generators as $f(\tilde T+a,\tilde U-a) = e^{a(\partial_{\tilde T}-\partial_{\tilde U})}f(\tilde T, \tilde U)$, where~ $\partial_{\tilde T}-\partial_{\tilde U}$ is the differential translation generator along that edge of the triangle. Similarly can be shown for the other edges of the triangle also.} $\widetilde T+\widetilde U=1$, and lies in the plane $\widetilde S = \textrm{constant}$. We show in the next section that the differential operator $D_S$ is associated with the \(S\)-channel. The same holds true for $D_T$ and $D_U$ as well.

Note in \eqref{4der}, the directional derivatives obey the identity $D_S+D_T+D_U=0$. Starting anywhere in the locus of the triangle, the three directional derivatives in \eqref{4der} translate it parallel to the edges. Since they trace the three oriented edges of the triangle, the point returns to its original position, thus leading to its closure. Given the homogeneity zero nature of the variables $\widetilde S$, the triangle is projective. Thus the geometry associated with the four point generating potential \eqref{4pot} is a projective triangle\footnote{Notice in figure \ref{fig:projective_triangle}, the edges of the projective triangle correspond only to the shifted-energy poles. The exchange poles however appear at vertices rather than as edges. Thus, the projective triangle gives a genuine positive geometry for the kinematic numerator (i.e. the structures without the exchange pole), and not for the full correlator.}.

\vspace{-12.5pt}\section{Gluon Four Point Function}\vspace{-10pt}
The gluon four point function in Grassmannian space
is given by \cite{Arundine:2026fbr}
\begin{align}\label{4ptYM}
F_4^{YM}=\frac{2\;\mathcal{H}_{4,YM}}{\Sigma}\Bigg( \frac{c_S}{S}\bigg(\frac{1}{\Delta_U}-\frac{1}{\Delta_T}\bigg)+\;\textrm{T, U-channels}\Bigg),
\end{align}
where $c_i$ is the color and $\mathcal{H}_{4,YM}$ is the helicity factor.

For the three point case, the generating potential produced the helicity-stripped correlator. However at four point, one obtains the helicity-stripped kinematic numerator (i.e. the structures without the exchange pole) associated with all the exchange channels. The full correlator is then obtained by attaching factorization poles to the kinematic numerators.

\vspace{5pt}\paragraph{\bf{Correlator as slopes along channel directions:}}
Consider the first derivative of the generating potential \eqref{4pot} with respect to $D_S$, which results in the helicity-stripped $S$-channel numerator \cite{bala2026color}
\begin{align}\label{ns}
n_S=D_S\;\Phi_4=\frac{2}{\Sigma}\bigg(\frac{1}{\Delta_U}-\frac{1}{\Delta_T}\bigg).
\end{align}
The helicity-stripped kinematic numerators for the other channels are: $n_T=D_T\Phi_4 \;\textrm{and} \;n_U=D_U\Phi_4$. Each numerator is the slope of the generating potential \eqref{4pot} along the their respective channel direction.

The full color-dressed four-gluon correlator is therefore obtained by attaching the kinematic numerators with their corresponding factorization poles, along with the helicity prefactor $\mathcal{H}_{4,YM}$ and the color factors $c_i$. 
\begin{align}\label{4ptYMFull}
F_4^{YM}=\mathcal{H}_{4,YM}\bigg( \frac{c_Sn_S}{S}+ \frac{c_Tn_T}{T}+ \frac{c_Un_U}{U}\bigg),
\end{align}
which is in agreement with \cite{Arundine:2026fbr,bala2026color}.

\vspace{5pt}\paragraph{\bf{Triangle geometry and kinematic Jacobi identity:}}
Closure of the three directions in the projective triangle figure \ref{fig:projective_triangle} implies
$ D_S+D_T+D_U=0$. The kinematic Jacobi identity \cite{bala2026color} is then immediate\footnote{A differential kinematic Jacobi identity was also obtained in Mellin-momentum space \cite{Mei:2023jkb}.}.
\begin{align}\label{KinJac}
n_S+n_T+n_U=(D_S+D_T+D_U)\Phi_4=0.
\end{align}
Thus closure of the projective triangle under directional derivatives \eqref{4der} provides a geometric realization of kinematic Jacobi identity.
\vspace{5pt}\paragraph{\bf{Projective rescaling and Kleiss--Kuijf identity:}}
Consider the color-ordered gluon correlator \cite{Arundine:2026fbr}
\begin{align}\label{colorordered}
 A(S,T,U)&=\frac{n_S}{S}-\frac{n_T}{T}
         =\frac1{ST}\left(
 \frac3\Sigma+\frac1{\Delta_U}
 -\frac1{\Delta_S}-\frac1{\Delta_T}\right),
 \end{align}
which can be rewritten as a simple derivative on $\Phi_4$
\begin{align}
 A(S,T,U)
 &=\frac{2}{ST}\,\partial_U\Phi_4.
\end{align}
Cyclic symmetry of $\Phi_4$ generates the other color orderings:
$A(T,U,S)=\frac{2}{TU}\partial_S\Phi_4$ and
$A(U,S,T)=\frac{2}{US}\partial_T\Phi_4$. Their sum leads to the Kleiss--Kuijf relation
\begin{align}
 A(S,T,U)+A(T,U,S)+A(U,S,T)=\frac{2}{STU}\,\hat E(\Phi)=0,
 \label{KK}
\end{align}
where $ \hat E=S\partial_S+T\partial_T+U\partial_U.$ The final equality follows from the zero homogeneity nature of the generating potential
\begin{align}
\Phi_4(\lambda S,\lambda T,\lambda U)=\Phi_4(S,T,U).
\end{align}
Thus, four point Kleiss--Kuijf identity \eqref{KK} is a direct consequence of zero homogeneity of the generating potential, which is encoded in the projective nature of the geometry. 

\vspace{-12.5pt}\section{Graviton Four Point Function via Geometry and Double Copy}\vspace{-10pt}
The graviton four point function in Grassmannian space is given by \cite{Bala:2026lvw,Arundine:2026fbr}
\begin{align}\label{4ptGR}
F_4^{GR}=\frac{8\;\mathcal{H}_{4,GR}}{\Sigma^2}\Bigg(\frac{1}{S}\bigg(\frac{1}{\Delta_U^2}+\frac{1}{\Delta_T^2}\bigg)+\textrm{T, U-channels}\Bigg).
\end{align}
We will obtain the same result using the generating potential \eqref{4pot} and its associated geometry \ref{fig:projective_triangle}.
For graviton, consider the second-order derivative of the generating potential \eqref{4pot}, which leads to the graviton $S$-channel numerator obtained via BCFW bridge \cite{Bala:2026lvw}
\begin{align}
D_S^2\;\Phi_4=\frac{8}{\Sigma^2}\bigg(\frac{1}{\Delta_U^2}+\frac{1}{\Delta_T^2}\bigg).
\end{align}
Note that the graviton numerator exhibits a differential double copy  $D_S^2\Phi_4$ in terms of the generating potential, as compared to the algebraic double copy obtained in \cite{bala2026color}. A simple double-copy $n_S^2$ of the Yang-Mills \eqref{ns} would have generated cross-terms $\sim\frac{1}{\Delta_T^2\Delta_U^2}$. Thus, the graviton probes the Hessian\footnote{On the triangle, the scalar potential is given by $2\Phi_4=-\log\widetilde S-\log\widetilde T-\log\widetilde U$
with $\widetilde S=1-\widetilde T-\widetilde U$. Choosing the coordinates $(\tilde T,\tilde U)$ its Hessian is $g_\triangle\equiv\partial_i\partial_j(2\Phi_4)=
\left(\begin{smallmatrix}
\tilde T^{-2}+\tilde S^{-2} & \tilde S^{-2}\\
\tilde S^{-2} & \tilde U^{-2}+\tilde S^{-2}
\end{smallmatrix}\right)$,
with $\det g_\triangle=(\tilde S^2+\tilde T^2+\tilde U^2)/
(\tilde S^2\tilde T^2\tilde U^2)>0$, so $g_\triangle$ is positive definite and defines a Hessian metric on the projective triangle. The graviton numerators are thus the diagonal components of the Hessian of $\Phi_4$.} of the same generating potential, whose slope gives the gluon and hence inherits the same origin. This renders a geometric interpretation to the graviton four point function.

The BCFW exchange parts of the correlator are then obtained by attaching the numerators of various channels to their corresponding factorization poles and the overall helicity prefactor $\mathcal{H}_{4,GR}=(\mathcal{H}_{4,YM})^2$, which leads to \eqref{4ptGR}.

\vspace{-12.5pt}\section{Spin-S Four Point Function}\vspace{-10pt}
Consider the four point function for equal spin-s fields with spin-s exchange. Performing the $s^{th}$ order derivative of the generating potential yields
\begin{align}\label{hsnum}
D_S^s\;\Phi_4=\frac{2^{2s-1}(s-1)!}{\Sigma^s}\bigg(\frac{1}{\Delta_U^s}+(-1)^s\frac{1}{\Delta_T^s}\bigg),
\end{align}
which agrees with the BCFW bridge \cite{Bala:2026lvw}. Thus, the spin-s numerator is $s^{th}$ differential copy of gluon numerator. 

For odd integer spin, the higher variation is antisymmetric between the two shifted-energy facets seen by a given channel direction. Cyclic sum of the numerators of the three channels makes the kinematic Jacobi identity manifest
\begin{align}
n_S^{(s)}+n_T^{(s)}+n_U^{(s)}=0,\quad \forall\; s\in2\mathbb{Z}_{\geq0}+1,
\end{align}
where we define the spin-s numerator for $i^{th}$ channel as
\begin{align}
n_i^{(s)}=\frac{1}{(s-1)!}D_i^s\;\Phi_4.
\end{align}
For even spin, the two facet contributions add rather than subtract, and no adjoint-type Jacobi identity follows.

The BCFW exchange part for the spin-s correlator is then simply
\begin{align}\label{4ptHS}
F_4^{(s)}=\mathcal{H}_4^{(s)}\sum_{i=S,T,U} \frac{n_i^{(s)}}{S_i},
\end{align}
which is true up to boundary terms whenever the BCFW deformation does not vanish at infinity.

Thus, the same projective triangle geometry that generates gluon and graviton correlators generates the higher-spin correlators.

\vspace{-12.5pt}\section{Connection to the BCFW bridge}\vspace{-10pt}
The differential operators \eqref{4der} acting on the generating potential \eqref{4pot} can be given a simple and very interesting interpretation in terms of the Grassmannian BCFW bridge \cite{Bala:2026lvw}. This connection also somewhat demystifies why a single scalar potential correctly produces all the correlators obtained directly from the BCFW bridge construction. Consider, for
example, the $|1,3\rangle$ deformation
\begin{equation}
 S(z)=S-zq,\quad
 T(z)=T+zq,\quad
 U(z)=U,
 \quad q=c_{23}c_{34},
\end{equation}
under which $\Sigma=S+T+U$ is unchanged. The deformed shifted-energy divisors become
\begin{equation}
 \Delta_S(z)=\Delta_S+2qz,\quad
 \Delta_T(z)=\Delta_T-2qz,\quad
 \Delta_U(z)=\Delta_U.
\end{equation}
The generating potential, consequently becomes
\begin{equation}
 \hat\Phi_4(z)
 =
\frac12\log\left(
 \frac{\Sigma^3}
 {(\Delta_S+2qz)(\Delta_T-2qz)\Delta_U}\right).
\end{equation}
Expanding around $z=0$ gives
\begin{align}
\hat\Phi_4(z) &= \hat\Phi_4(0)
 +z\frac{\partial}{\partial z}\hat\Phi_4(0)
 +\frac{z^2}{2!}
   \frac{\partial^2}{\partial z^2}\hat\Phi_4
 +\mathcal O(z^3),\notag \\
 &= \Phi_4
 +z\,\frac{\Sigma q}{2}\,
   D_U\Phi_4
 +\frac{z^2}{2}
   \left(\frac{\Sigma q}{2}\right)^2
   D_U^2\Phi_4
 +\mathcal O(z^3),
\end{align}
where using the chain rule, the $z$-expansion has the directional derivatives \eqref{4der}
\begin{equation}
 D_U
 =\frac{2}{\Sigma}
   \left(\partial_T-\partial_S\right)=\frac{4}{\Sigma^2}(\partial_{\widetilde S}-\partial_{\widetilde T}).
\end{equation}
The first order expansion of the scalar potential along the BCFW deformation parameter $z$
produces the gluon kinematic numerator for the U-channel, while the second order expansion produces the corresponding graviton numerator. Higher powers in the expansion have the same interpretation. More generally, the $z^s$ coefficient is proportional to
$D_U^{\,s}\Phi_4$. The other channels follow from the other bridge deformations $|1,4\rangle$ and $|1,2\rangle$.  In this sense, the same scalar potential generates the spin-one and spin-two structures by successive derivatives along
Grassmannian BCFW flows.

\vspace{-12.5pt}\section{Discussion $\And$  Outlook}\vspace{-10pt}
Our starting point for this paper was the following question: which object produces an exact differential double copy relation between gluon and graviton? Demanding such a relation resulted in the scalar generating potential $\Phi$. This raises the question of whether all double copy relations discussed in the literature can be organized by a scalar function analogous to $\Phi$.

A natural follow up would be to generalize the generating potential $\Phi$ for higher point functions and its corresponding geometric perspective using appropriate criteria \cite{XYZ:2026new4}. The scalar potential can be a natural setup for constructing a Parke-Taylor like formula for (A)dS, which would be far more simple than dealing with gauge fields. Just like the case of three and four point function, once we obtain the scalar potential for the higher point function, one can obtain graviton or any spin-s correlator by taking appropriate derivatives. 

One of the remarkable results we have obtained is that there exists a common geometric origin for gluons, gravitons and higher-spin for cosmological correlators. This is in contrast to \cite{Bala:2026lvw,Arundine:2026qqg}, where only the geometry for gluons is discussed. In flat space, there has been attempts to give a geometric interpretation with the graviton \cite{Trnka:2020gravituhedron}, but a complete geometric explanation remains open. We speculate that a scalar function akin to the generating potential might be a natural object that leads to a geometric picture for graviton scattering, even in flat space.

\vspace{5pt}\vspace{5pt}\paragraph{\bf{Possible connection to Associahedron and Kinematic flow:}}
The projective triangle associated with auxiliary potential \(\Phi_4\) also has an interesting connection to the four point associahedron \cite{ArkaniHamed:2017associahedron}. For example, one-dimensional boundary,  the edge \(\Delta_U=0\) of the triangle, for which \(U=S+T\) may be parametrized by
\[
x\equiv \frac{S}{S+T},\qquad 1-x=\frac{T}{S+T}.
\]
The endpoints \(x=0,1\) corresponds respectively to the \(S=0\) and \(T=0\) factorization channels, and the induced canonical form is given by
\[
\operatorname{Res}_{\Delta_U=0}\Omega_4
=
d\log\frac{x}{1-x}
=
\frac{dx}{x(1-x)}.
\]
This implies each edge of the projective triangle is identified with a four point associahedra. Let us emphasize that this is only a boundary correspondence: the full triangle, and in particular the full gluon correlator, is not identified with the associahedron.

It is interesting to note that there is also a suggestive connection with kinematic flow \cite{ArkaniHamed:2023flow}. The auxiliary potential satisfies
\[
d\Phi_4
=
\frac32\,d\log\Sigma
-\frac12\sum_{I=S,T,U}d\log\Delta_I.
\]
The kinematic-flow systems are in general described in terms of   logarithmic structure
\[
d\mathbf I=\sum_\alpha M_\alpha\,d\log L_\alpha\,\mathbf I .
\]
Both therefore, organize their singularities  through \(d\log\) forms. However, this resemblance is currently structural. We don't claim any well defined  equivalence. To summarize, the kinematic flow describes a closed system of master functions, whereas here \(\Phi_4\) is an auxiliary generating potential whose derivatives produce spinning numerators.
\vspace{7.5 pt}
\textbf{Acknowledgment:}
SJ would like to thank organizer and participants of the Holostrings III held at IIT Kanpur
where some aspects of results in this paper were discussed. AB acknowledges a UGC-JRF fellowship. We would like to acknowledge our debt to the people of India for their steady support of research in the basic sciences.
\bibliography{biblio}
\end{document}